# The citation impact of social sciences and humanities upon patentable technology

Felix de Moya-Anegon[1], Carmen Lopez-Illescas[2], Vicente Guerrero-Bote[3], Henk F. Moed[4]

[1]*felix.moya@scimago.es*
SCImago Group, Madrid, Spain

[2]*carmlopz@gmail.com*
University Complutense of Madrid. Information Science Faculty. Dept. Information and Library Science, SCImago Group, Spain

[3]*guerrero@unex.es*
SCImago Group, Dept. Information and Communication, University of Extremadura, Badajoz, Spain

[4] *henk.moed@uniroma1.it*
Sapienza University of Rome, Italy

## Abstract

This paper examines the citation impact of papers published in scientific-scholarly journals upon patentable technology, as reflected in examiner- or inventor-given references in granted patents. It analyses data created by SCImago Research Group, linking PATSTAT's scientific non-patent references (SNPRs) to source documents indexed in Scopus. The frequency of patent citations to journal papers is calculated per discipline, year, institutional sector, journal subject category, and for "top" journals. PATSTAT/Scopus-based statistics are compared to those derived from Web of Science/USPTO linkage. A detailed assessment is presented of the technological impact of research publications in social sciences and humanities (SSH). Several subject fields perform well in terms of the number of citations from patents, especially Library & Information Science, Language & Linguistics, Education, and Law, but many of the most cited journals find themselves in the interface between SSH and biomedical or natural sciences. Analyses of the titles of citing patents and cited papers are presented that shed light upon the cognitive content of patent citations. It is proposed to develop more advanced indicators of citation impact of papers upon patents, and ways to combine citation counts with citation content and context analysis.

## 1. Introduction

### *Citation analysis and the science-technology interface*

The relationships between science and technology constitutes one of the most important topics in quantitative science and technology studies. Citation analysis is one of the key methodologies that are used to study these relationships. Figure 1 presents a schematic overview of the application of citation analysis in the study of the science-technology interface. At the science side, citations in scientific articles to other scientific papers are used to analyse the cognitive structure of science, collaboration and knowledge flows among authors, and to assess the contribution scientific entities such as individual scholars, groups and departments made to scientific progress.

At the technology side, citations from one patent to another provide partial indications of the economic, technical or strategic value of patents, and of knowledge flows and collaboration networks among inventors. Citations in the scientific literature to patents mark the influence of patents in their role of scientific publications upon the scientific literature. Finally, citations given in patents to the scientific literature are used to study the influence of scientific-scholarly work upon technological development. It is this perspective that plays a key role in the current paper.

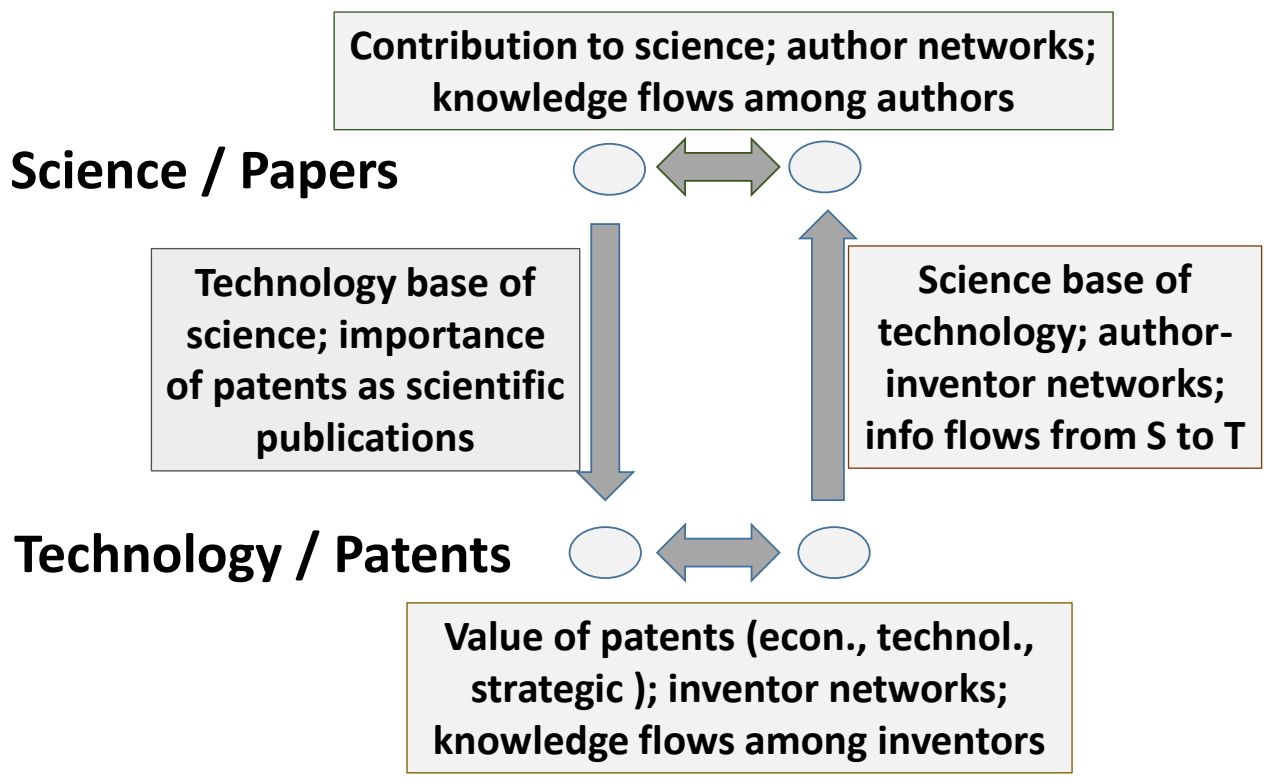

Figure 1. Citation links in the science-technology interface

*Brief literature review on the analysis of scientific non-patent references in patents*

Anthony van Raan (2017) gives an excellent overview of main developments in the use of patent citations to the scientific literature, starting with the work of Francis Narin and co-workers who explored measures of the "science intensity" of technological fields. They showed already in the 1990s how major inventions patented by industrial firms at the US Patent Office depend upon publicly funded basic research (e.g., Narin, Hamilton, & Olivastro, 1997). Van Raan used the acronym SNPRs to indicate Scientific Non-Patent References in patents. He concluded that "only a small minority of about 3-4 percent of publications covered by the Web of Science or Scopus are cited by patents. However, for publications based on university-industry collaboration the number of SNPRs is considerably higher, namely around 15%" (Van Raan, 2017, p. 13). The studies discussed by Van Raan are based on the analysis of non-patent references in USPTO (US Patent and Trademark Office), linked with the Web of Science or its predecessors (mainly the Science Citation Index).

Several studies revealed that not only the distribution of SNPRs on the "cited side" among *target* WoS papers is skewed, but also, on the "citing side", the distribution of SNPRs among *source* patents. Of course, these manifestations of skewness of citations on the citing and cited side have also been observed in the citation analysis of scientific papers (e.g., Price, 1965), but the skewness is for patent-to-paper citations much stronger than it is for paper-to-paper citations. Other studies observed a national patent citation bias: patents submitted by applicants from a particular country showed a preference for citing research papers by authors located in the same country. Other papers found a positive correlation between a country's technological performance on the one hand, and its scientific strength on the other, and provided evidence that in *emerging fields* of technology the number of SNPRs in patents is higher than it is in other fields (Van Looy et al., 2006).

In his review, Van Raan underlined that the number of SNPRs in patents, and the probability that a scientific publication may be used as an SNPR, depends on a series of factors, including the stage of development of a technological field; the distribution of SNPRs among inventors and examiners; characteristics of the patent office and the applicant firms; and differences in the economic value of patents. He concluded that "SNPRs indeed form a bridge between science and technology, but more in a broader sense, i.e. at a macro-level such as the "science intensity" of technological fields or the science-technology interaction at the level of countries" (Van Raan, 2017, p. 22). Defining the time lag between a scientific breakthrough and an invention as "the time lapse between the publication year of a paper and the year this paper is cited in a patent", Van Raan pointed out that large differences appear to exist in time lags between technological fields. He also underlined that "the SNPRs may represent important recent scientific research but this research on its turn may be based on even more important, earlier breakthrough work, not cited in the patent but perhaps cited in the SNPRs."

*Measuring the technological impact of scientific-scholarly subject fields*

At the technology side, in the analysis of linkages between patent citations and scientific-scholarly papers, measures are calculated of the science intensity, science base or science linkage of (patentable) technology. Francis Narin and co-workers defined science linkage as "a measure of the extent to which a company's technology builds upon cutting edge scientific research. It is calculated on the basis of the average number of references on a company's patents to scientific papers, as distinct from references to previous patents. Companies whose patents cite a large number of scientific papers are assumed to be working closely with the latest scientific developments" (Narin, Breitzman & Thomas, 2004).

From the science side, patent citations can be used to calculate indicators of the technological impact of scientific-scholarly work. Such indicators aim to capture the extra-scientific or 'societal' impact of research. For instance, Halevi and Moed (2012) examined the impact of research published in library science journals upon technology as reflected in SNPRs, using the LexisNexis product TotalPatent™ (TotalPatent, n.d.) linked with Scopus. A good overview of methodological approaches to statistical patent analysis is given in Hinze & Schmoch (2004).

*Aim of the paper and research questions*

The aim of the *first* part of this paper is to give a comprehensive overview of the frequency at which patents processed in PATSTAT cite Scopus source articles. The research questions addressed in the first part are as follows.

- What is the percentage of journal papers cited in patents and the average number of patent citations per journal paper in the various scientific-scholarly disciplines?
- How does the frequency of patent-to-paper citations change over the years during the time period 2008-2017, and how does it vary between institutional/economic sectors, research disciplines and journals?
- How do the outcomes obtained in the PATSAT/Scopus database compare with those based on linkages between USPTO non-patent references with Web of Science, and presented in Van Raan's review?

The outcomes provide a statistical background to the analyses presented in the *second* part of this paper, which focuses on social sciences and humanities (SSH). This part partly follows an analysis model applied in an earlier paper by Halevi and Moed (2012) on the technological impact of library science. The latter study found that research papers in library science had a considerable impact as reflected in patent citations. The current paper addresses whether such impact can also be found in other subject fields in social sciences and humanities (SSH). It presents a series of in-depth case studies of SSH subject fields and journals. This part of the current paper presents a first exploration of the links between citing patents and cited SSH papers and their meaning. More detailed analyses will be presented in future publications, including a paper by Guerrero-Bote, Moed, & Moya-Anegon (n.d.). The following research questions are addressed.

- How often are articles published in SSH journals cited in patents? Do the obtained numbers support the hypothesis that SSH fields, similar to science and biomedical fields, have an influence on technology as well?
- How do patent-to-paper citation counts vary across SSH subject fields and journals?
- What types of knowledge transfer may be reflected in patent citations to SSH papers? Typical examples are presented of citing patent and cited paper titles in selected subject fields
- Which are the affilations of the applicants of patents citing journal papers in selected SSH subject fields?

*A note on terminology*

In the current article, scientific documents published in journals and other sources processed for Scopus are denoted as *articles* or *papers*. Scopus contains many document types. In the current paper only *articles*, *reviews, mini-reviews* and *conference papers* are counted. These types are denoted below as *citable documents*. All these types are denoted as *articles*. The largest part of articles in Scopus is published in scientific-scholarly journals. But Scopus also processes conference proceedings and books as sources of citations. For the sake of brevity, and to avoid the ambiguous term *sources*, all source entities are denoted as *journals*. Citations made in patents to Scopus articles are denoted as *patent-to-paper citations*, and citations in one article indexed in Scopus to another Scopus-indexed article as *paper-to-paper citations*.

## 2. Data collection and handling

PATSTAT, "EPO worldwide PATENT STATistical Database", is a global patent database created by the European Patent Office (EPO, 2018), published for the first time in 2008, to help patent statistical research at the request of a working group on patent statistics led by the Organization for Economic Cooperation and Development (OECD). Other members of this working group are: World Intellectual Property Organization (WIPO), Japan Patent Office (JPO), United States Patent and Trademark Office (USPTO), Korea Intellectual Property Office (KIPO), National Science Foundation of the United States (NSF) and European Commission (CE).

As main advantages over other databases such as NBER (USA) or IIP (Japan), PATSTAT has worldwide coverage, includes more types of information and contains some auxiliary products that solve some of its problems, which has made it a de facto standard (Kang and Tarasconi, 2016). Its disadvantages are its orientation towards Europe (data from national offices are exchanged with the EPO on the basis of agreements that change over time and may leave gaps) and its orientation to the review process (data that are not vital in the process of the patent examination has a lower quality).

PATSTAT is a relational database. It can be purchased on a DVD to be installed on a local computer or online, and can be consulted using SQL (De Rassenfosse, Dernis and Boedt, 2014).

The EPO publishes two annual editions of PATSTAT, Spring and Autumn. The 2018 Spring Edition of PATSTAT (PATSTAT - Spring Edition of 2018) is a snapshot of the data present in DOCDB EPO, a global bibliographic database that includes data from more than 90 patent offices around the world, and the global database of legal information of INPADOC EPO. State, taken in the fifth week of 2018.

One of the PATSTAT tables includes the references to the non-patent literature. This table contains the full non-patent references, which do not follow a fixed format, and are not always complete. The table also contains a series of related data fields, but in many cases the values in these fields are missing. In a combined automated and manual approach, the records in this table were matched one by one against the source documents indexed in Scopus. This work was carried out by SCImago Research Group These authors have designed a procedure divided in four phases (Guerrero-Bote, Sánchez-Jiménez & Moya-Anegón, 2019):

1. Data preprocessing: Preparation of data to facilitate and streamline subsequent processes. The most important actions: unify records; locate patterns corresponding to DOIs; assign publication years; normalize lexical variants and eliminate special characters; locate possible elements of the reference: first author, title, source; generate an inverted index with extracted terms.
2. Pre-selection of candidate couples. With the previous data of the pre-processing phase, we have 9x1014 possible pairs formed by a NPL reference of PATSTAT and a reference of Scopus. Due to the lack of standardization, a direct comparison is necessary and that is impossible to address in such a large number of couples. For that reason, this phase aims to reduce that number, reduce that number to a sufficiently large number to minimize the possibility of a real couple being left out. To this end, a series of rules are used that are applied in the form of SQL statements in the data obtained from the previous phase.
3. Automatic evaluation of the candidate couples. The objective of this phase is to assign a score that allows to select for each NPL reference the Scopus reference that probably refers to the same document. For this purpose, a series of routines have been designed that look for the most important elements of the Scopus reference in the record containing the non-patent reference. The overall score is obtained by the product of the scores obtained for each element of the reference (by way of probability).
4. Human validation. An NPL reference may not have a Scopus candidate reference, it may have one or it may have several. Logically, if any of the candidates corresponds to the NPL reference, this should be the highest score obtained, but it is possible none of the assigned ones was valid. For this reason, a manual validation is necessary. To this end, an application has been developed that allows the cooperation of many people in human validation. For more information about the data handling, the reader is referred to Guerrero-Bote, Moed, & Moya-Anegon, F. (n.d.).

A patent family can be defined as "a set of patents taken in various countries to protect a single invention (when a first application in a country – the priority – is then extended to other offices)." In other words, a patent family is "the same invention disclosed by a common inventor(s) and patented in more than one country (“Patent families”, n.d.). One of the problems in patent citation analysis is that there may be substantial differences between members of the same family as regards the non-patent references they may contain. This is especially the case for the *examiner*-given references, as patents of a family tend to pass a different evaluation process in each office. In some cases, this process is faster and in others slower, and some incorporate more non-patent literature references than others. To avoid these differences, SCImago Research Group has retrospectively assigned all non-patent references in the various members of a family to each patent in that family. In this way, when a patent is granted, it incorporates all scientific non-patent references in its *entire family*.

## 3. Results (all disciplines)

*Overall results and analysis by institutional sector*

Table 1 presents a breakdown of papers and patent citations by institutional sector of the (cited) papers. Data relate to the time period 2008-2017. For papers the variable year relates to the publication year, while for patents it refers to the *filing* year. A paper may be assigned to multiple institutional sectors, if it results from a collaboration between authors active in institutions located in different sectors (e.g., a university-company collaboration). Therefore, the total number of paper-sector assignments exceeds the number of papers, by some 15 per cent. For all sectors combined, the share of papers cited in patents is 3.2 % if double counts due to these multiple assignments are included, and 2.7 % otherwise. The largest percentage of papers from a particular sector cited in patents, relative to the total number of papers assigned to that sector, is obtained by the *private* sector (7.9 %), followed by the *health* sector (4.2 %). The share of private sector papers relative to the total number of papers in all sectors is only 2.6 %, but its share relative to the total number of received patent citations is 11 %, similar to that for the government sector.

**Table 1. Number of papers and patent citations by papers' institutional sector (time period 2008-2017)**

| *Papers Institutional Sector* | *Papers* | | *Citations in Patents to Papers* | | *# Papers Cited in Patents* | | *% Papers Cited in Patents* |
|---|---|---|---|---|---|---|---|
| | *N* | *%* | *N* | *%* | *N* | *%* | |
| Higher Education | 18,534,000 | 68.1% | 4,129,000 | 53.0% | 527,000 | 60.1% | 2.8 |
| Government | 4,137,000 | 15.2% | 930,000 | 11.9% | 134,000 | 15.3% | 3.2 |
| Health | 3,565,000 | 13.1% | 1,811,000 | 23.2% | 150,000 | 17.1% | 4.2 |
| Private | 719,000 | 2.6% | 854,000 | 11.0% | 57,000 | 6.5% | 7.9 |
| Other | 250,000 | 0.9% | 66,000 | 0.8% | 8,000 | 0.9% | 3.2 |
| Total paper-sector assignments | 27,205,000 | 100.0% | 7,790,000 | 100.0% | 876,000 | 100.0% | 3.2 |
| Total unique papers (excl. 'double counts') | 23,511,000 | | 5,351,000 | | 628,000 | | 2.7 |

Van Raan (2017) indicated in his review an overall percentage of papers cited in patents of 3-4, obtained in studies that were based on USPTO and WoS. In the current study, based on Scopus and PATSTAT, this percentage is somewhat lower. As noted above, Table 1 also shows that the percentage of cited papers in patents (last column) is for *paper-sector assignments* (semi-last row, 3.2 %) somewhat larger than that for total number of *unique* papers (bottom row, 2.7%). This means that multi-sector papers tend to attract somewhat more patent citations than single-sector publications do. Data for collaborative papers between the public and the private sector are not available in the current study. Therefore, van Raan's conclusion (Van Raan, 2017) that papers co-published between the public and private sector have a relatively large percentage of papers cited in patents, cannot be directly validated.

*Trends in annual patent-to-paper citation rates*

Table 2 presents the annual trend in the number and percentage of papers cited in patents during the 10-year period 2008-2017. A 5-year citation window is used. This means that only citations (in patents) are counted to publications published during the *five* years preceding the filing year of a patent. In the column headers journal papers are denoted as *docs*.

**Table 2. Annual trend in the number and percentage of papers cited in patents between 2008-2017**

| Year | Nr Journals (sourceids) with at least 1 citable doc in Scopus in past 5 years | Nr journals receiving at least 1 patent citation to citable docs in past 5 yrs | % journals receiving at least 1 patent citation to citable docs in past 5 yrs | Number of citable docs in 5 preceding years in Scopus | Number of citable docs from past 5 years cited in at least one patent | % citable docs from past 5 years cited in at least one patent | Total patent citations to citable docs from past 5 years | Number of patent citations per citable doc (5yrs) | Total citing patent families to citable docs from past 5 years | Number of citing patent families per citable doc (5yrs) |
|---|---|---|---|---|---|---|---|---|---|---|
| 2008 | 23,820 | 8,424 | 35.4% | 7,847,445 | 176,321 | 2.25% | 598,989 | 0.076 | 248,725 | 0.032 |
| 2009 | 25,368 | 8,779 | 34.6% | 8,325,768 | 182,296 | 2.19% | 600,541 | 0.072 | 259,075 | 0.031 |
| 2010 | 27,240 | 9,349 | 34.3% | 8,825,158 | 188,379 | 2.13% | 607,578 | 0.069 | 266,009 | 0.030 |
| 2011 | 28,716 | 9,856 | 34.3% | 9,291,259 | 191,230 | 2.06% | 609,970 | 0.066 | 272,490 | 0.029 |
| 2012 | 30,094 | 10,305 | 34.2% | 9,783,094 | 197,467 | 2.02% | 639,670 | 0.065 | 278,183 | 0.028 |
| 2013 | 30,758 | 10,727 | 34.9% | 10,250,199 | 196,259 | 1.91% | 600,011 | 0.059 | 272,744 | 0.027 |
| 2014 | 31,203 | 10,607 | 34.0% | 10,722,918 | 185,699 | 1.73% | 575,083 | 0.054 | 252,377 | 0.024 |
| 2015 | 31,890 | 9,940 | 31.2% | 11,203,655 | 155,557 | 1.39% | 496,650 | 0.044 | 207,536 | 0.019 |
| 2016 | 32,313 | 9,015 | 27.9% | 11,575,432 | 119,278 | 1.03% | 382,423 | 0.033 | 157,430 | 0.014 |
| 2017 | 32,669 | 7,697 | 23.6% | 11,879,030 | 79,459 | 0.67% | 245,469 | 0.021 | 101,388 | 0.009 |

Table 1 above reveals an overall percentage of papers cited in patents of around 3 per cent. The percentages in Table 2 are lower, namely 1-2 %. The difference is explained by the fact that the percentages presented in Tables 1 and 2 are based on different citation windows: Table 1 includes citations to papers from the total time period (1996-2017), while Table 2 applies a 5-year citation window.

The total number of patent citations to citable docs (5-year window) shows after 2012 a rapid *decline*, despite the fact the total number of citable docs increased during these years. To understand this, one should note that the analyses presented in the current paper relate to patents granted before 2018, and that the patent examination process may take several years, so that there may be a considerable time delay between a patent's filing year and the year it is granted. The number of (source) patents filed in a given (filing) year *and* granted before 2018 declines as the filing year becomes more recent, because of this delay in granting of filed patents. For instance, the major part of patents filed in 2017 will be granted in 2018 or later; as a result, the citations in these patents to papers are *not* included in the counts presented in Table 2. Since according to Table 2 the maximum number of patent citations to citable documents (published in the past 5 years) is obtained for the filing year *2012*, the following sub-sections in this analyse data for this year.

*Analyses by subject category*

Figure 2 compares for each of about 300 Scopus journal categories an indicator similar to the standard *journal* impact factor (JIF) based on paper-to-paper citations with one based on *patent-to-paper* citations. The horizontal axis displays the mean value of a journal's number of *paper-to-paper* citations per article across all journals covering a particular Scopus journal category, applying a three-year window, counting citations in 2012 to citable documents published in the three preceding years 2009-2011.

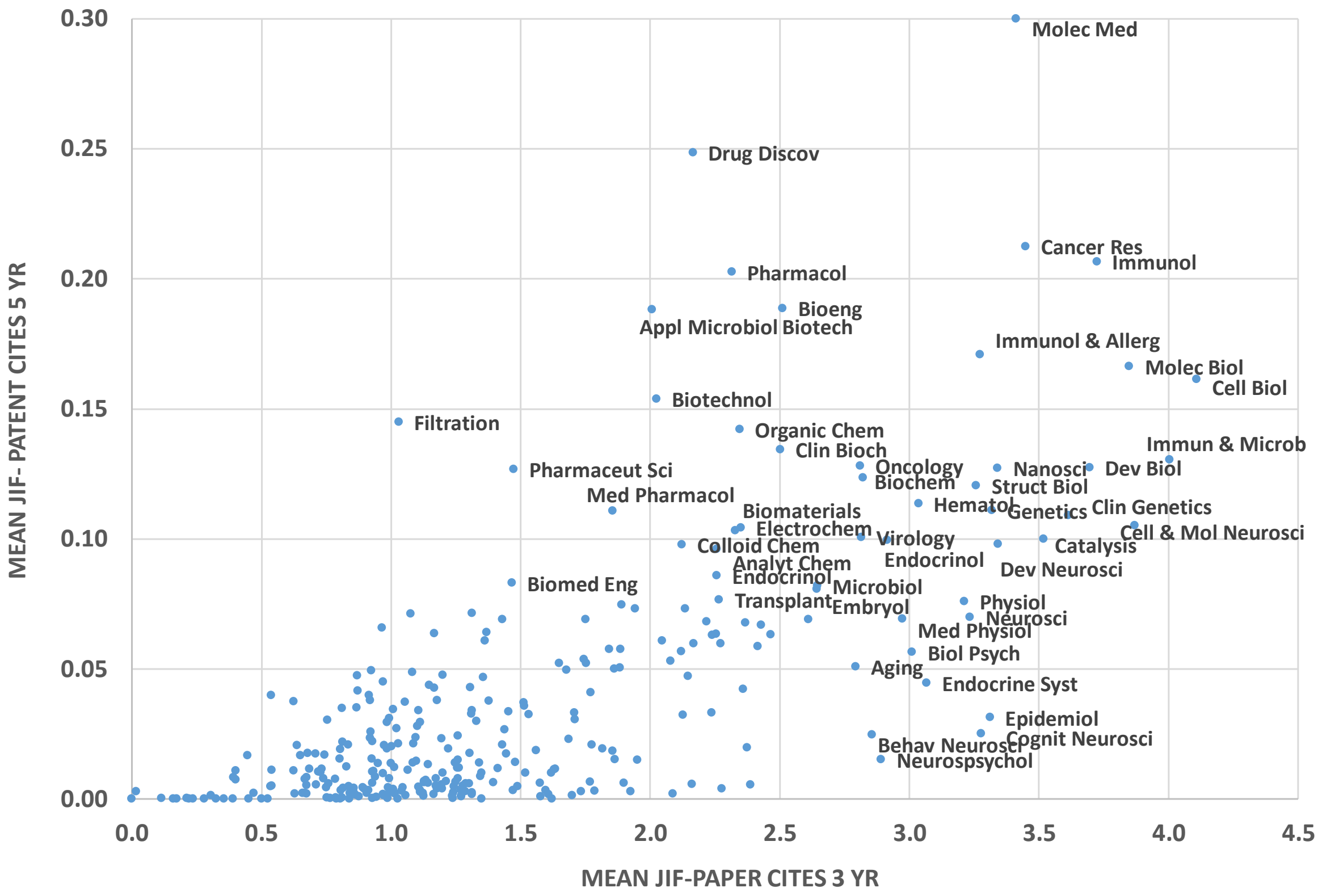

**Figure 2. Mean journal impact factors based on paper citations and patent citations, per journal category, and for the (first filing or citation) year 2012.**

The vertical axis gives a similar measure but now based on *patent-to-paper* citations, and applying a 5-year rather than a 3-year window. Biomedical categories tend to show on average the largest patent citation-based JIFs. *Relative* differences between categories hardly change if one analyses *patent family* citations instead of *patent* citations. The mean citation rates for these two types of citations per category show a very strong linear correlation: R-square is 0.97.

*Analyses by journal*

The paper-to-paper- and patent-to-paper-based journal impact factor, displayed in Figure 2, are also calculated in columns 5 and 7 in Table 3 below, but here at the level of *individual* journals instead of journal *categories*. Table 3 lists the 10 journals with the largest impact score based on patent citations for a single year: 2012. As indicated in Table 2 above, at the level of *all journals in all fields combined*, and for the year 2012, the average number of patent citations per citable document is 0.065. The level of the patent citation impact of the *top journals* in Table 3 ranges between 0.9 and 3.8. Table 3 also shows in column 6 a citation impact measure based on (citing) *patent families* rather than *citing patents*. Roughly speaking, the patent family-based scores are one-third to one-half of those based on patent citations.

**Table 3. The 10 journals with the largest percentage of 2007-2011 papers cited in 2012 patents**

| Rank | Journal Title | Nr citable docs in 2007-2011 | % citable docs 2007-2011 cited in 2012- patents to | JIF based on number of patent citations and 5-year window | JIF based on number of patent *family* citations and 5-year window | JIF based on paper citations and 3-year window | SJR (Scimago Journal Rank) 2012 based on paper citations |
|---|---|---|---|---|---|---|---|
| 1 | Annual Review of Immunology | 119 | 42.9 | 2.21 | 0.95 | 57.5 | 38.09 |
| 2 | mAbs (covering Antibody R&D) | 196 | 37.2 | 2.45 | 0.78 | 4.8 | 1.40 |
| 3 | Advanced Drug Delivery Reviews | 576 | 35.6 | 2.33 | 0.97 | 15.4 | 4.50 |
| 4 | Progress in Polymer Science | 214 | 35.5 | 1.60 | 0.95 | 31.1 | 10.00 |
| 5 | Pharmacological Reviews | 107 | 31.8 | 1.81 | 0.57 | 24.8 | 10.67 |
| 6 | Journal of Medicinal Chemistry | 3,907 | 31.3 | 1.68 | 0.52 | 5.9 | 2.34 |
| 7 | Proc Int. Symp. on Computer Architecture | 210 | 31.0 | 0.93 | 0.41 | 10.2 | 2.33 |
| 8 | Nature Biotechnology | 988 | 30.9 | 3.85 | 1.33 | 19.1 | 10.87 |
| 9 | Metabolic Engineering | 260 | 30.4 | 1.44 | 0.61 | 7.5 | 2.99 |
| 10 | Molecular Therapy | 1,194 | 30.3 | 1.12 | 0.36 | 7.9 | 3.23 |

## 4. Results for Social Sciences and Humanities

*Results per subject category*

An analysis per subject category in Social Scienses and Humanities (SSH) is complicated by many questionable or erroneous assignments in Scopus of journals to subject categories. For instance, the journals *Science* and *Annals of the New York Academy of Sciences* are fully assigned to the Arts and Humanities category *History and Philosophy of Science*, although the number of papers in these journals about history and philosophy of science is very limited. Including all papers in these two journals into this category gives a totally distorted picture of this field. There are also assignments that seem completely erroneous. For instance, *Journal of Fluorescence*, covering papers on an established spectroscopic technique, is linked with the Social Science categories *Miscellaneous, Social and Political Science*, and *Law*. *Journal of the Acoustical Society of America* is assigned to Arts and Humanities *(Miscellaneous*), and *Physiology and Behavior* and *European Child and Adolescent Psychiatry* to the category *Philosophy*.

In most categories, the distribution of patent citations is very skew, and the journals with the largest score tend to have a strong orientation towards science and (bio)medical fields. This is illustrated in Table 4, that gives for categories with more than 20 patent citations the total number of patent citations in the filing year 2012 to papers published in the 5 preceding years, and a list of at most 3 journals with the largest number of patent citations. The questionable assignments mentioned above are *not* included in this table. Also, conference proceedings are not included in the shortlist of most cited journals, but their citations are included in the total counts for a category.

**Table 4: The journals most frequently cited in patents filed in 2012, per journal category**

| Category and journal | | Nr Pat Cites | Category and journal | | Nr Pat Cites |
|---|---|---|---|---|---|
| *Archeology (arts and humanities)* | | *17* | *Geography, Planning and Development* | | *200* |
| | Journal of Cultural Heritage | 7 | | Building and Environment | 51 |
| | Radiocarbon | 6 | | ISPRS Journal of Photogrammetry and Remote Sensing | 29 |
| | Journal of Archaeological Science | 3 | | International Archives of the Photogrammetry, Remote Sensing and Spatial Information Sciences - ISPR | 21 |
| *History and Philosophy of Science* | | *26* | *Health (social science)* | | *101* |
| | Ars Pharmaceutica | 9 | | MMWR. Morbidity and mortality weekly report | 24 |
| | Social Science and Medicine | 5 | | Alcohol | 14 |
| | Philosophy, Ethics, and Humanities in Medicine | 5 | | Trauma, Violence and Abuse | 12 |
| *Language and Linguistics* | | *341* | *Human Factors and Ergonomics* | | *153* |
| | Computational Linguistics | 32 | | International Journal of Human Computer Studies | 41 |
| | Speech Communication | 26 | | Color Research and Application | 18 |
| | Artificial Intelligence | 9 | | Journal of Physiological Anthropology | 17 |
| *Music* | | *271* | *Law* | | *325* |
| | AES: Journal of the Audio Engineering Society | 135 | | IEEE Security and Privacy | 136 |
| | IEEE Signal Processing Magazine | 111 | | Computer Standards and Interfaces | 25 |
| | Acta Acustica united with Acustica | 6 | | Electricity Journal | 21 |
| *Anthropology* | | *54* | *Library and Information Sciences* | | *613* |
| | Journal of Physiological Anthropology | 17 | | Journal of Chemical Information and Modeling | 223 |
| | Collegium Antropologicum | 16 | | IEEE Transactions on Information Theory | 200 |
| | Journal of Cultural Heritage | 7 | | Information Processing and Management | 39 |
| *Communication* | | *107* | *Safety Research* | | *191* |
| | International Journal of Digital Multimedia Broadcasting | 59 | | Therapeutics and Clinical Risk Management | 147 |
| | Speech Communication | 26 | | Journal of Occupational Medicine and Toxicology | 12 |
| | Journal of Advertising Research | 3 | | Environmental Biosafety Research | 8 |
| *E-Learning* | | *145* | *Sociology and Political Science* | | *33* |
| | IEEE Transactions on Learning Technologies | 16 | | Educational Technology and Society | 5 |
| | Computers and Education | 13 | | Social Networks | 5 |
| | Journal of Digital Information | 7 | | Population and Development Review | 4 |
| *Education* | | *321* | *Transportation* | | *122* |
| | Chemical Research in Chinese Universities | 50 | | IET Intelligent Transport Systems | 17 |
| | International Journal of Human Computer Studies | 41 | | Journal of Air Transport Management | 4 |
| | Measurement: Journal of the International Measurement Confederation | 37 | | Transportation Research Part E: Logistics and Transportation Review | 4 |

*Results for selected SSH journals*

While Halevi and Moed (2012) obtained evidence of an economical and technological impact of the field *library science*, the current study aims to explore whether such influence can also be found in other SSH subject fields. To that end, apart from library science, three additional, typical journal categories from SSH were chosen: *History and Philosophy of Science*, *Music*, and *Education*. Also *Library and Information Science* was included in order to compare results with those obtained by Halevi and Moed (2012). Data presented in Table 5 relate to patents in PATSTAT filed during the time period 2008-2017 (and granted before 2018), and citations to Scopus articles published in the five years preceding the patent's filing year. Table 5 presents for 12 journals in these four SSH subject fields typical examples of titles of citing patents and cited papers. With the purpose of gaining insight into the specific subjects of the citing patents and cited papers, word clouds are presented using the Worditout software (http://worditout.com) for all four fields: Library & Information Science (Figures 3a and 3b), Education (figures 4a and 4b), History and Philosophy of Science (figures 5a and 5b), and Music (figures 6a and 6b). These maps give a first impression of the contents of citing patents and cited papers. The prominent position of generic, methodological terms such as 'Method' and 'System', especially in the maps based on citing patents is itself informative: it marks the importance of findings and approaches from the scientific literature for the development of new, patented methodologies, some with a broad scope. Moreover, also non generic terms are displayed in the maps.

According to the results presented by Halevi and Moed (2012), library science papers cited in patents were mostly featuring library information and customer management systems together with classification and indexing methodologies, while the patents citing these articles were found to feature mainly online commerce applications. Also the current analysis illustrates the important role played by these subjects in the titles of citing patents as well as in cited papers. In the world clouds for Library & Information Science (Figures 3a and 3b) the importance of web based technology is clearly reflected in the font size (indicating their frequency of occurrence) of words such as *web*, *google*, *machine*, *technologies*, *computing*, *engine*, *link*, *databases*, *automated*, *multimedia*, *browsing*, *network*, *site or XML*. However, and not surprisingly, the more important words displayed in the word clouds for papers and patents in the journal *Scientometrics* are related to information and evaluation. Among these words one finds: *measures, search*, *analysis*, *similarity*, *data*, *document*, *information*, *content*, *knowledge*, *journal*, *performance*, *survey*, *table*, *evaluating*, *factorizing*, *matrix*, *graph*, *weighted*, *comparison*, *results*, *bibliometric*, *citation*, *clusters*, *quality*, *quantitative*, *mapping* or *ranking*.

**Table 5. Examples of patent citations to selected SSH journals**

| *Field / Journal* | *Citing patent title (Example)* | *Cited paper title (Example)* |
|---|---|---|
| ***Library & Information Science*** | | |
| Journal of Information Science (56/15) | SYSTEM AND METHOD FOR ANNOTATING DOCUMENTS | Usage patterns of collaborative tagging systems |
| | METHOD AND DEVICE FOR ENRICHING A CONTENT DEFINED BY A TIMELINE AND BY A CHRONOLOGICAL TEXT DESCRIPTION | Aspect-based sentiment analysis of movie reviews on discussion boards |
| | SYSTEMS AND METHODS OF HANDLING INTERNET SPIDERS | Web robot detection in the scholarly information environment |
| Journal of Informetrics (6/3) | INCORPORATING LEXICON KNOWLEDGE INTO SVM LEARNING TO IMPROVE SENTIMENT CLASSIFICATION | Sentiment analysis: A combined approach |
| | SYSTEMS AND METHODS FOR RANKING NODES OF A GRAPH USING RANDOM PARAMETERS | Finding scientific gems with Google's PageRank algorithm |
| | SYSTEM AND METHOD FOR ANALYZING AND CATEGORIZING TEXT | Applying social bookmarking data to evaluate journal usage |
| Scientometrics (18/14) | APPARATUS AND METHOD FOR DETERMINING STAGE USING TECHNOLOGY LIFECYCLE | Anticipating technological breakthroughs: Using bibliographic coupling to explore the nanotubes paradigm |
| | SYSTEM METHOD AND PROGRAM TO TEST A WEB SITE | 'Mini small worlds' of shortest link paths crossing domain boundaries in an academic Web space |
| | SOCIAL NETWORK MODEL FOR SEMANTIC PROCESSING | A quantitative analysis of indicators of scientific performance |
| ***History & Philosophy of Science*** | | |
| British Journal for the Philosophy of Science (15/2) | DIAGNOSABILITY SYSTEM: FLOOD CONTROL | Causes and explanations: A structural-model approach. Part I: Causes |
| IEEE Annals of the History of Computing (17/5) | DESKTOP STREAM-BASED INFORMATION MANAGEMENT SYSTEM | Alan Kay: Transforming the computer into a communication medium |
| | METHOD AND SYSTEM FOR PROCESSING EMAIL ATTACHMENTS | The technical development of internet email |
| | AMBIENT BACKSCATTER TRANCEIVERS APPARATUSES SYSTEMS AND METHODS FOR COMMUNICATING USING BACKSCATTER OF AMBIENT RF SIGNALS | Implications of historical trends in the electrical efficiency of computing |
| Philosophy, Ethics, and Humanities in Medicine (54/2) | COMPOSITION AND METHODS OF USE FOR TREATMENT OF MAMMALIAN DISEASES | Are animal models predictive for humans? |
| | NOVEL BIOMARKERS | Rethinking psychiatry with OMICS science in the age of personalized P5 medicine: Ready for psychiatome? |
| ***Music*** | | |
| Computer Music Journal (58/12) | APPARATUS AND METHOD FOR ENHANCED SPATIAL AUDIO OBJECT CODING | The spatial sound description interchange format: Principles, specification, and examples |
| | METHOD AND SYSTEM FOR EXTRACTING TEMPO INFORMATION OF AUDIO SIGNAL FROM AN ENCODED BIT-STREAM AND ESTIMATING PERCEPTUALLY SALIENT TEMPO OF AUDIO SIGNAL | Complexity-scalable beat detection with MP3 audio bitstreams |
| | SYSTEM AND METHOD FOR PERFORMING AUTOMATIC AUDIO PRODUCTION USING SEMANTIC DATA | An offline, automatic mixing method for live music, incorporating multiple sources, loudspeakers, and room effects |
| Music Perception (26/8) | PERSONALIZED AUDITORY-SOMATOSENSORY STIMULATION TO TREAT TINNITUS | Listening to filtered music as a treatment option for tinnitus: A review |
| | SYSTEMS METHODS AND MEDIA FOR IDENTIFYING MATCHING AUDIO | Perceptual smoothness of tempo in expressively performed music |
| | METHODS AND DEVICES FOR TREATING HYPERTENSION | Music and autonomic nervous system (dys)function |
| Organised Sound (13/2) | APPARATUS AND METHOD FOR EFFICIENT OBJECT METADATA CODING MULTI-CHANNEL AUDIO SIGNALS | Object-based audio reproduction and the audio scene description format |
| | WEARABLE SOUND | Imposing a networked vibrotactile communication system for improvisational suggestion |
| Psychology of Music (18/3) | SIDE EFFECT AMELIORATING COMBINATION THERAPEUTIC PRODUCTS AND SYSTEMS | Exposure to music and cognitive performance: tests of children and adults |
| | A MEDIA PLAYER | Toward a better understanding of the relation between music preference, listening behavior, and personality |
| | SYSTEMS AND TECHNIQUES FOR IDENTIFYING AND EXPLOITING RELATIONSHIPS BETWEEN MEDIA CONSUMPTION AND HEALTH | Listening to sad music in adverse situations: How music selection strategies relate to self-regulatory goals, listening effects, and mood enhancement |
| ***Education*** | | |
| Computers and Education (74/29) | DISEASE THERAPY GAME TECHNOLOGY | Exploring the potential of computer and video games for health and physical education: A literature review |
| | PERSONALIZING ERROR MESSAGES BASED ON USER LEARNING STYLES | E-Learning personalization based on hybrid recommendation strategy and learning style identification |
| | INDIVIDUAL LEARNING DEVICE AND METHOD BASED ON RADIO COMMUNICATION NETWORK | The design and evaluation of a computerized adaptive test on mobile devices |
| Proceedings - Frontiers in Education Conference (28/11) | HIERARCHICAL STATE MACHINE GENERATION FOR INTERACTION MANAGEMENT USING GOAL SPECIFICATIONS | Invited panel - Engineering technology education in the era of globalization |
| * | CLOUD DESKTOP SYSTEM WITH MULTI-TOUCH CAPABILITIES | Web-based tools to sustain the motivation of students in distance education |
| | TAGGING METHOD | Web editing module for tagging metadata of the Fedora commons repository learning objects under DRD and LOM standards |

Numbers between parentheses give the number of patent citations and the number of cited papers per journal

The current results also coincide with those obtained by Halevi and Moed as regards the prominent role of indexing and classification methodologies in library science. What it is important to notice in the current picture is the appearance of numerous and highlighted words related to indexing and classification connected to the web based technologies. In the word clouds we find: *classification*, *social*, *tag*, *tagging*, *bookmarking*, *folksonomy*, *collaborative*, *community*, *index*, *ontology* or *semantic*. The technological influence of the research carried out in indexing and classification in relation to the social and semantic web is even more evident in the papers and patents' titles: *collaborative tagging*, *sentiment analysis*, *web robot detection*, *Google's PageRank*, *social bookmarking*, *social network*, *semantic processing*, among others.

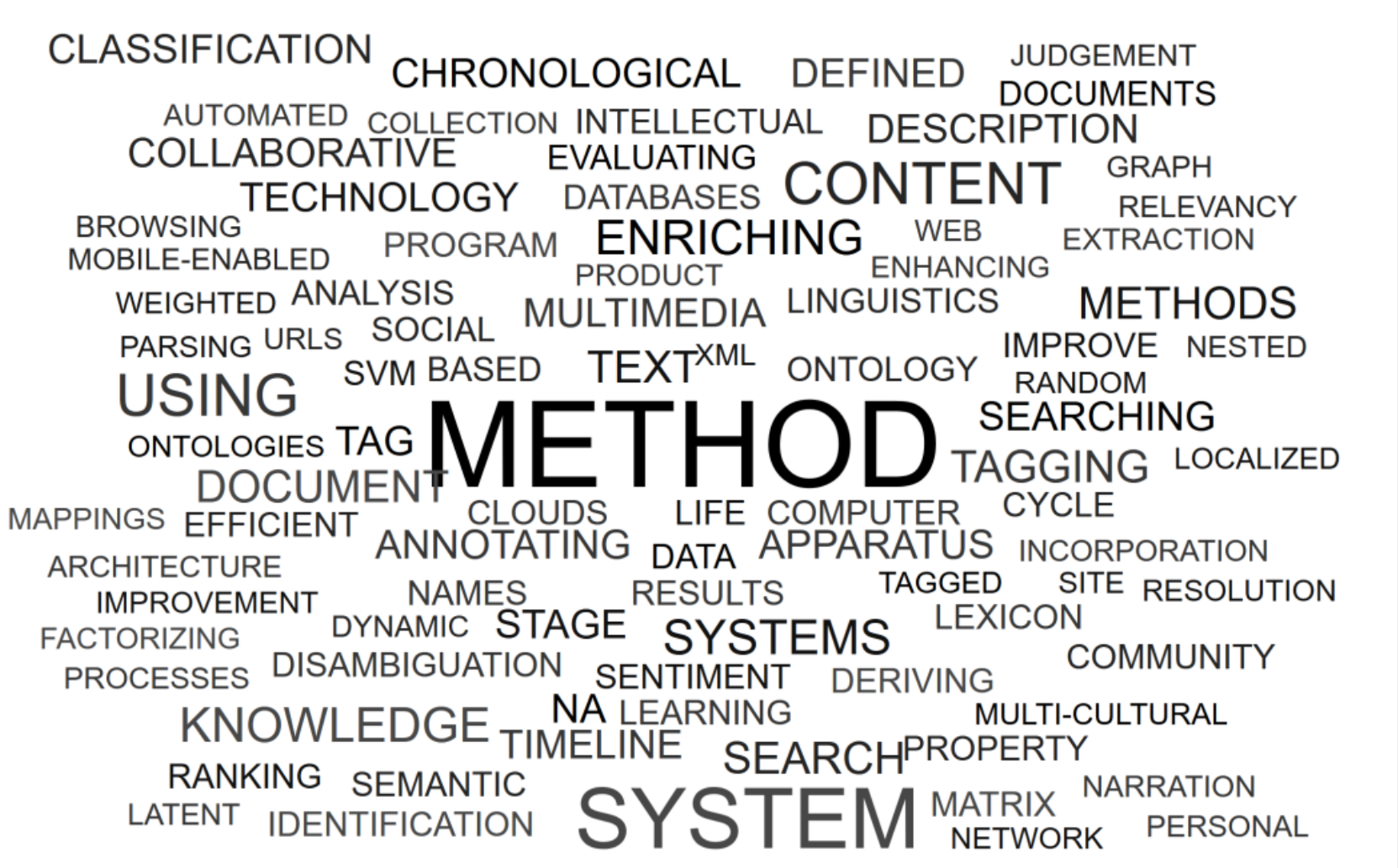

**Figure 3a. Word Clouds for Library & Information Science based on citing patent titles**

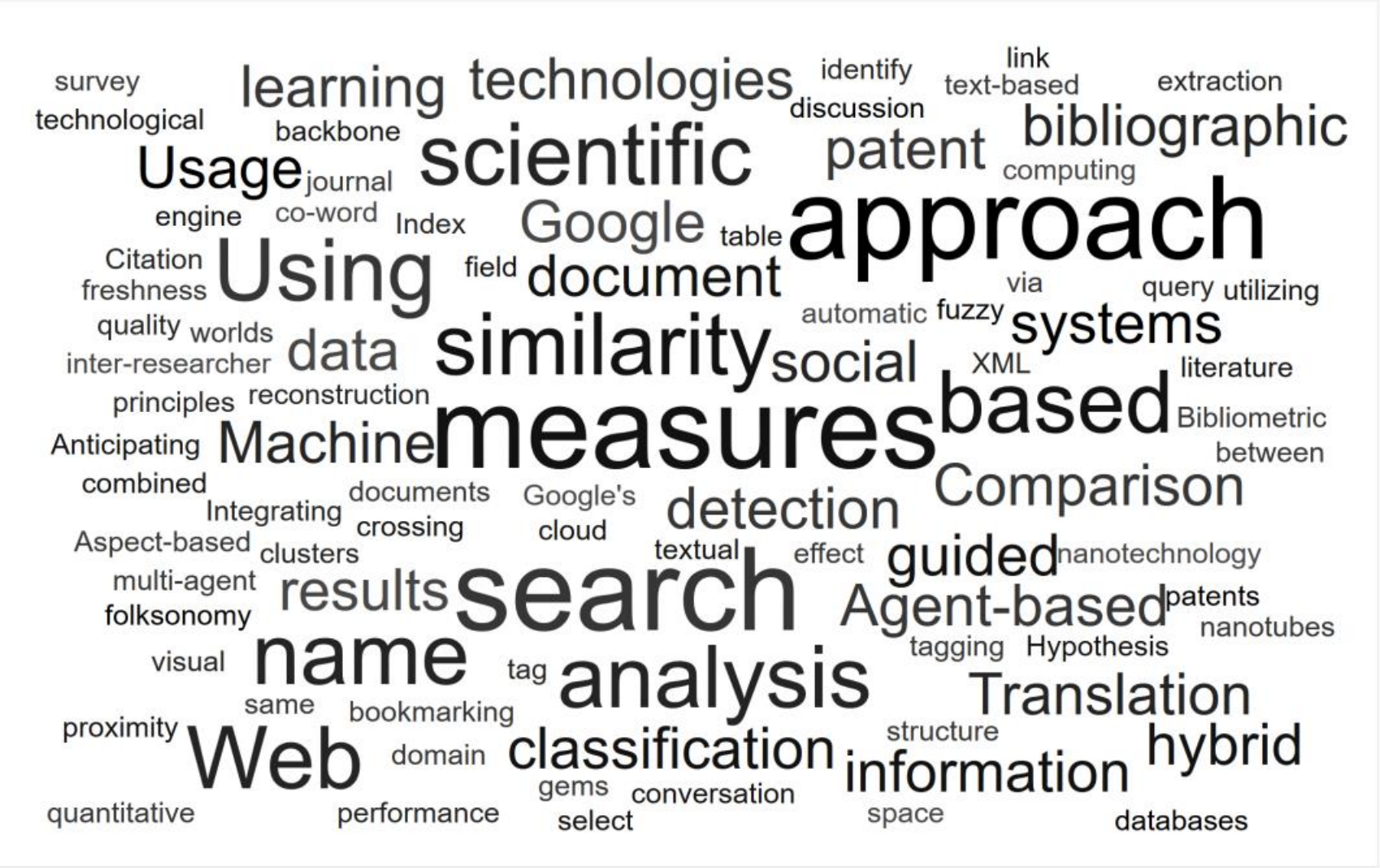

**Figure 3b. Word Clouds for Library & Information Science based on cited paper titles**

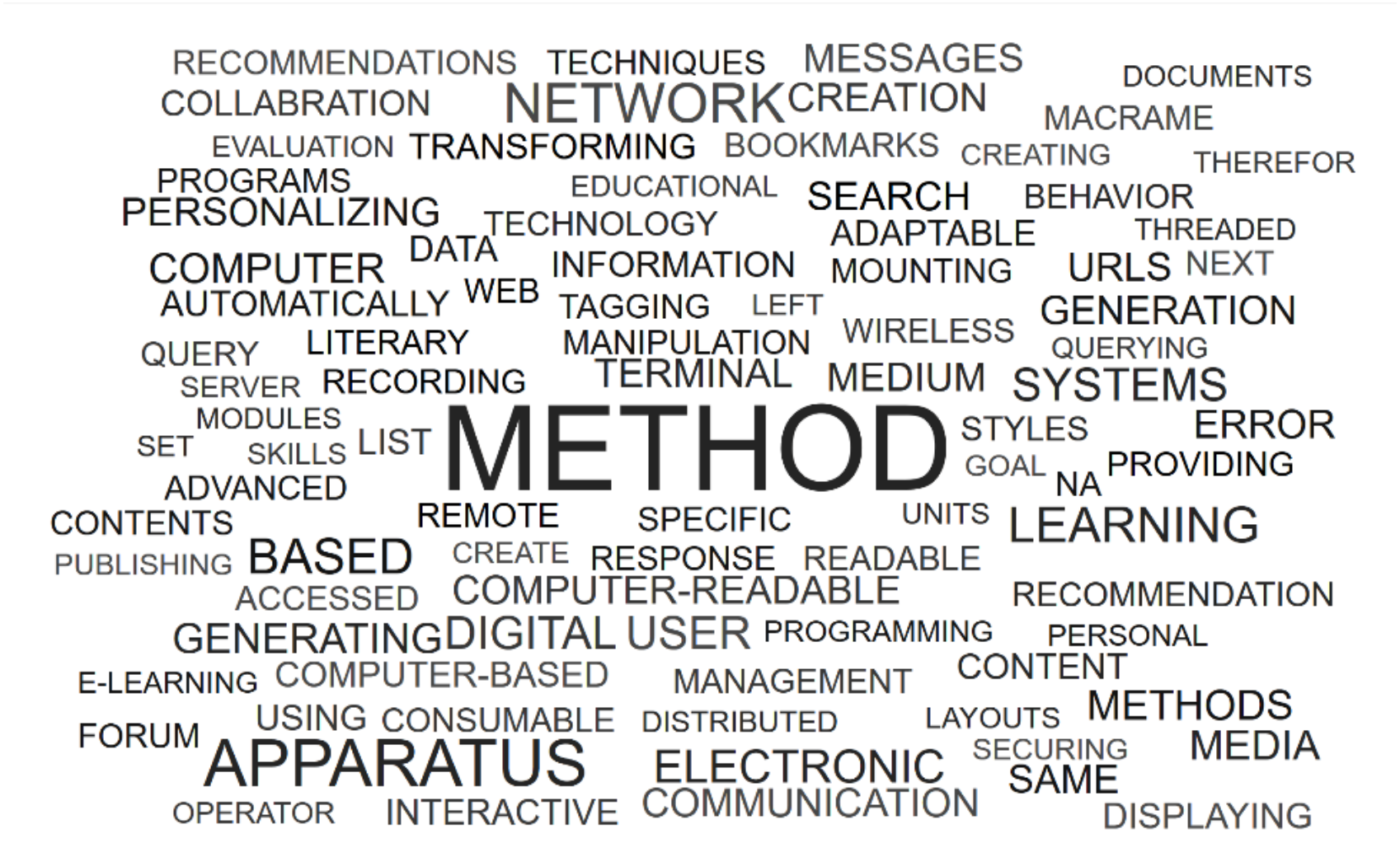

**Figure 4a. Word Clouds for Education based on citing patent titles**

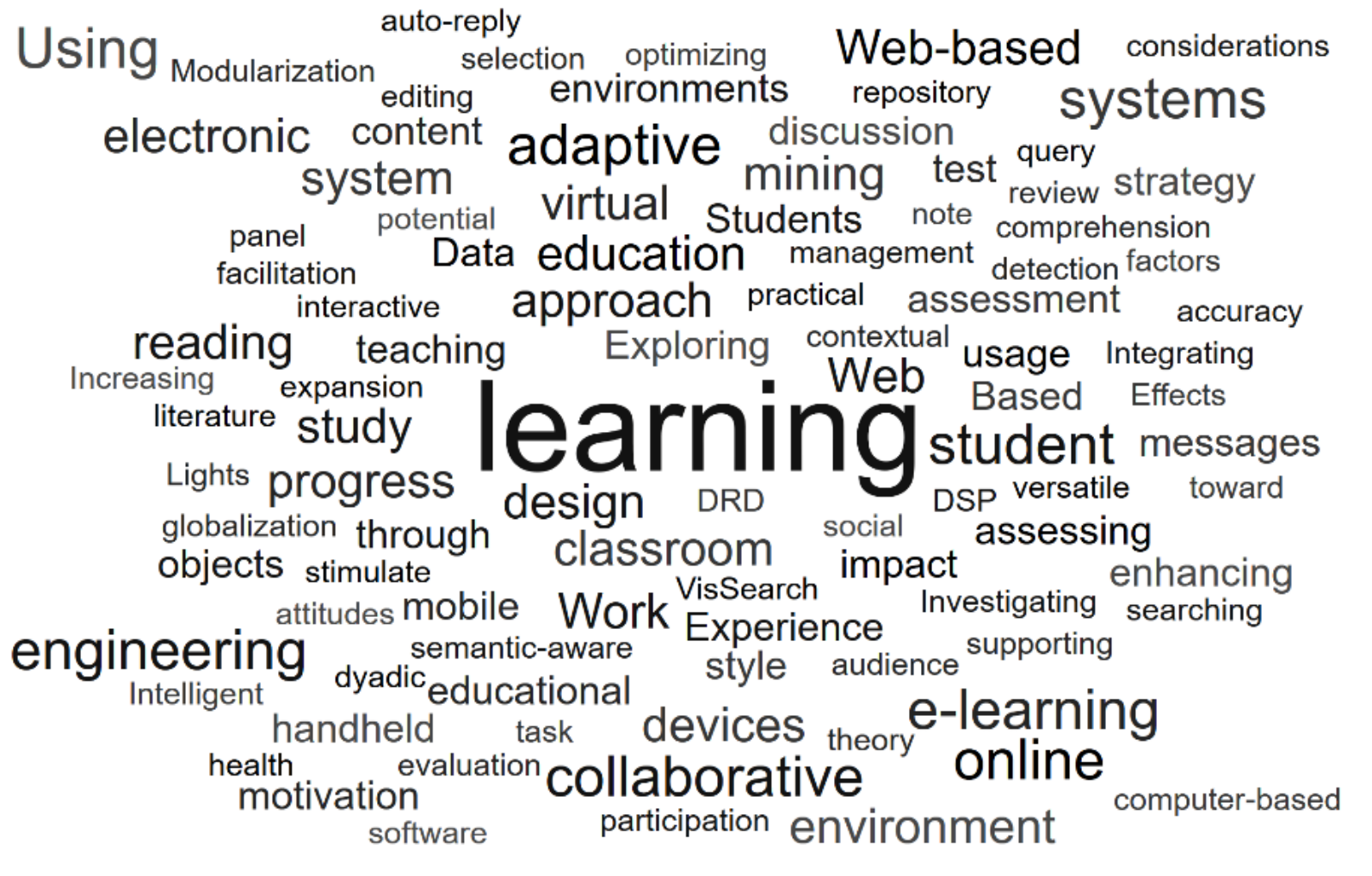

**Figure 4b. Word Clouds for Education based on cited paper titles**

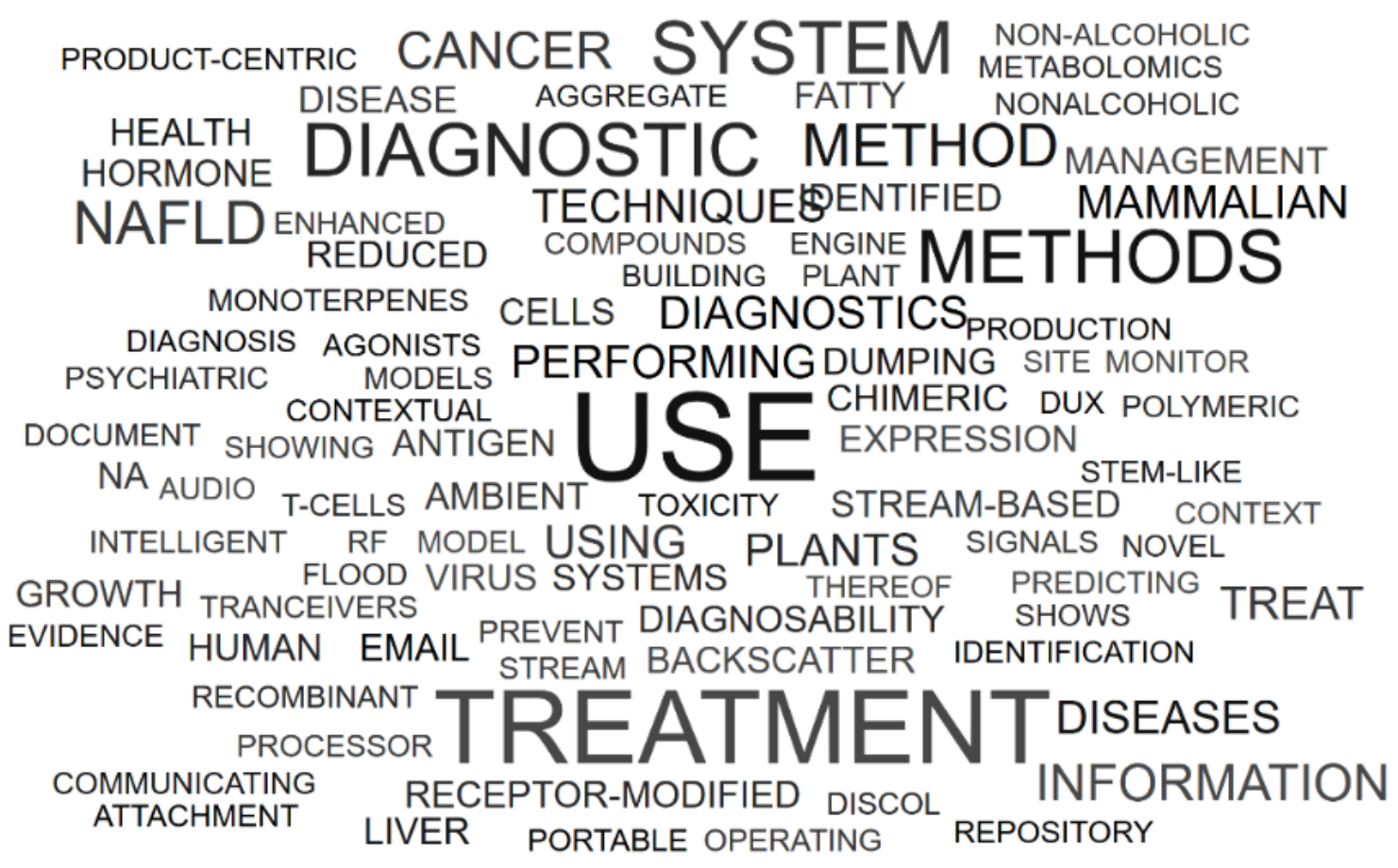

**Figure 5a. Word Clouds for History & Philosophy of Science based on citing patent titles**

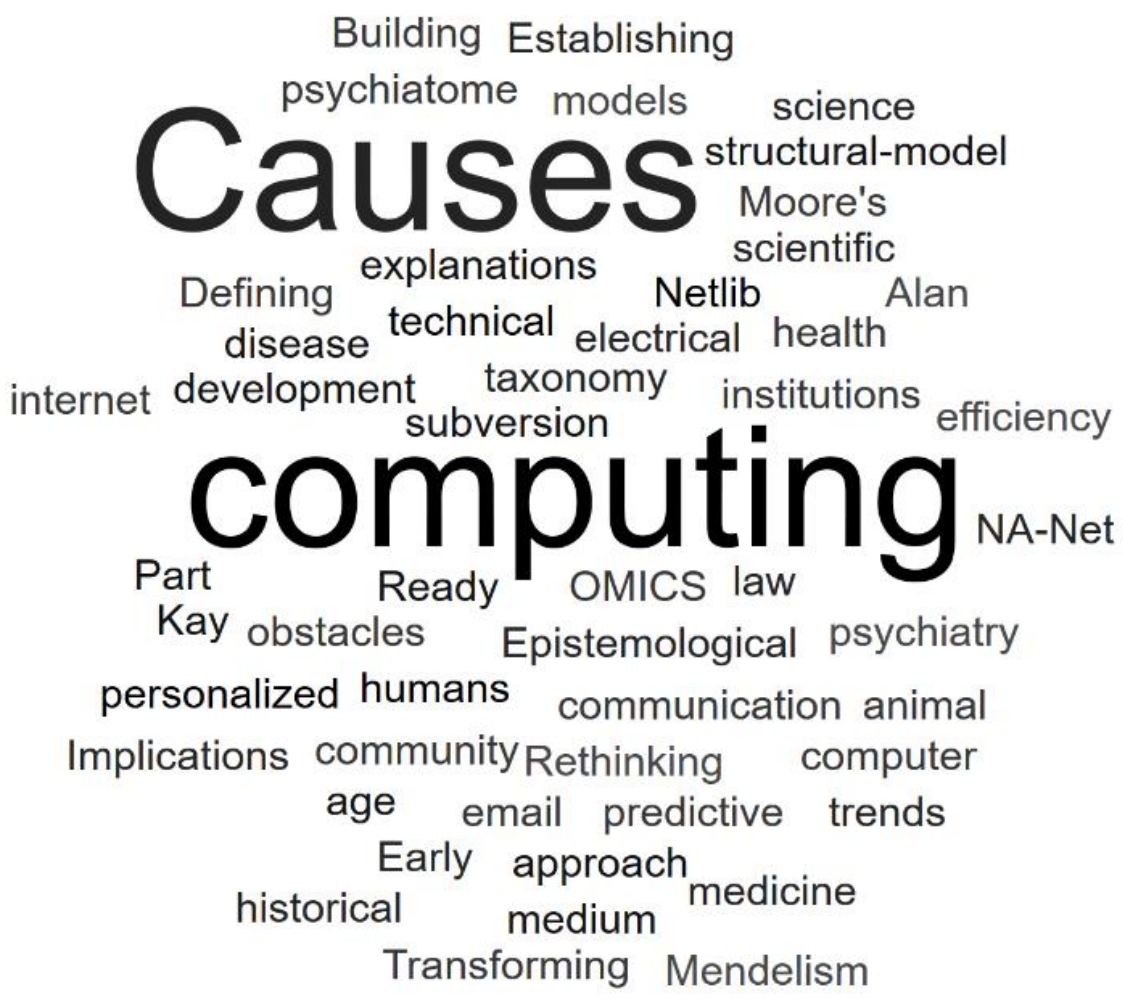

**Figure 5b. Word Clouds for History & Philosophy of Science based on cited paper titles**

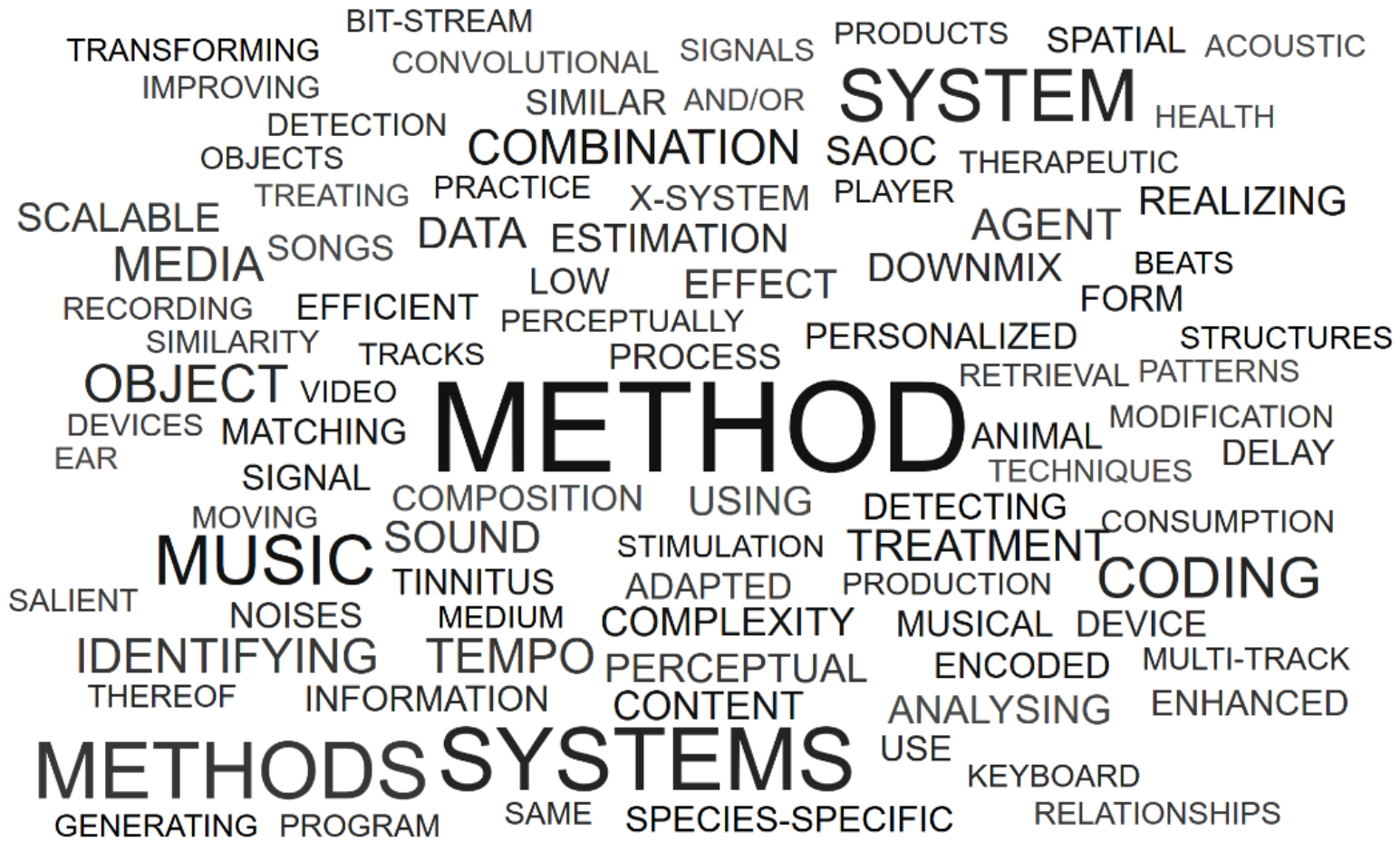

**Figure 6a. Word Clouds for Music based on citing patent titles**

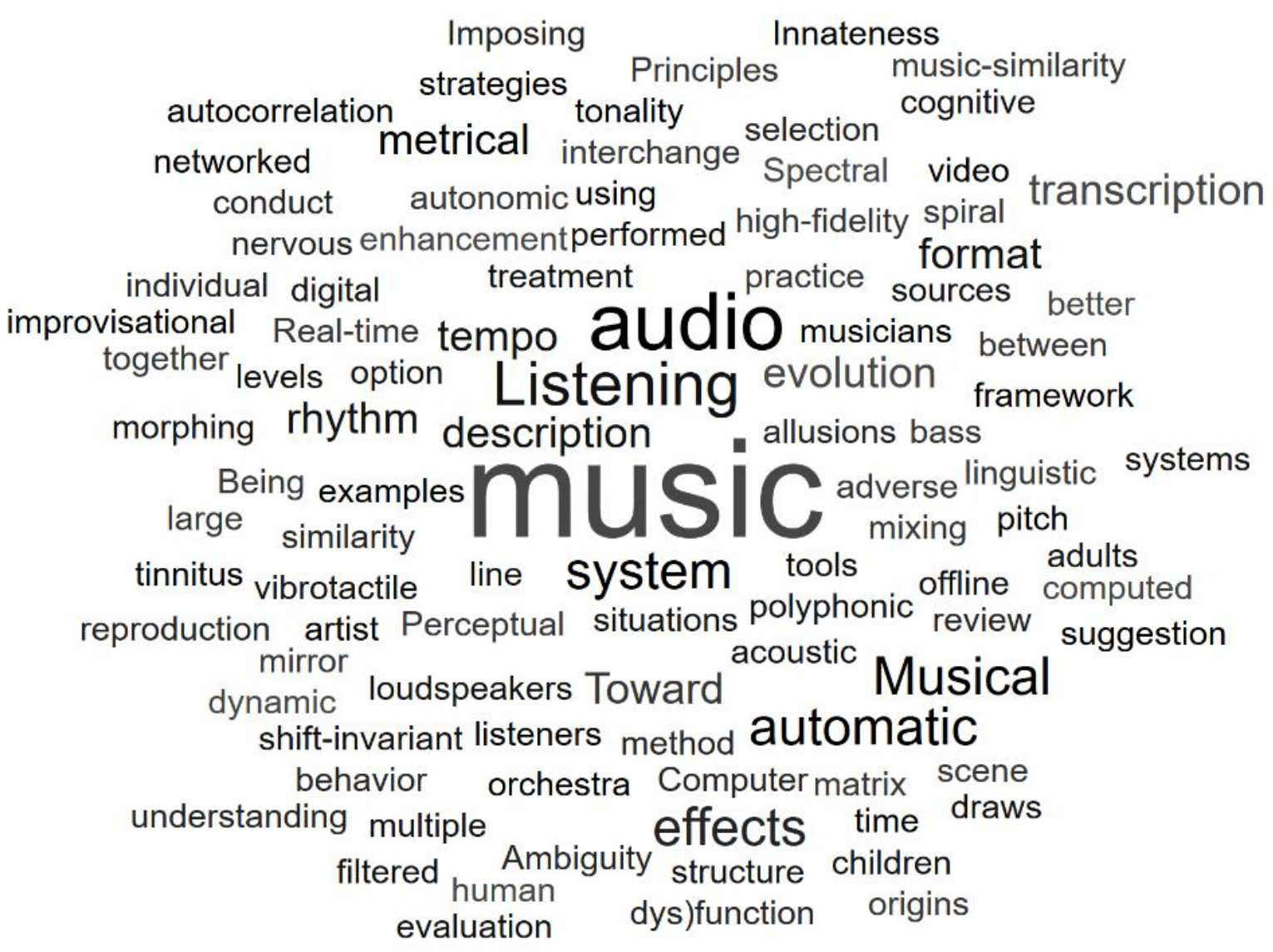

**Figure 6b. Word Clouds for Music based on citing patent titles**

The shift in emphasis from the Web 1.0 to the challenging Web 4.0, and from document sharing to data sharing reflects that library science is on the basis of the information science-technology interaction. The relationship between the development of library science and innovation in Web technology is not new, but these results confirm the increasing importance of the technological impact of the subject field, given the expanding role of these technologies in redefining today's society. The upcoming trends in the fields of information technologies reveal the world as a globalized highly intelligent information space, where the individual needs will be socially customized.

This combination of the social and personalization aspect is also present in the results obtained from the other studied SSH categories. Words related to both customization and social issues are present as well in the word clouds and in the patent and papers' titles. We find the words *collaborative*, *social*, *personalizing*, *personal*, *community*, *personalized* or *individual* appear with a high frequency. In the other three analysed fields the research cited in patents were mostly featuring Web and electronic systems. For Education we find the words: *e-learning*, *online*, *electronic*, *virtual*, *web*, *web-based*, *computer-based*, *software*, *network*, *digital*, *computer-readable*, *urls*, *automatically*, *wireless*, *programming*, *remote* and *server*. In music, *automatic* is one of the most highlighted words, also *computed*, *computer*, *digital*, *offline*. In History the most visible word is *computing*. Computing is one of the most common connection to other fields found in the SSH research cited in patents, but not the only one. We find connection to Health in all categories except for Library & Information Science. In Education, Music and even more in History & Philosophy of Science we find words standing out such as *health*, *therapeutic*, *disease*, *medicine*, *diagnostic*, *cancer*, *diagnosability*, *antigen*, *cells* or *virus*.

Table 6 gives a list of the most important patent assignees in the four subject fields. These lists do not merely include large technology firms, but also an investment company (Invention Science Fund I), a university (Univ Pennsylvania) and a research organization (Fraunhofer).

**Table 6. Most important assignees per research field**

| *Library & Information Science* | | *History & Philosophy of Science* | | *Music* | | *Education* | |
|---|---|---|---|---|---|---|---|
| *Assignee* | *# Patents* | *Assignee* | *# Patents* | *Assignee* | *# Patents* | *Assignee* | *# Patents* |
| IBM | 13 | ORACLE | 15 | FRAUNHOFER | 9 | IBM | 5 |
| PALO ALTO RES. CENTER (Xerox) | 7 | BOEHRINGER INGELHEIM | 6 | THE INVENTION SCIENCE FUND I | 8 | MICROSOFT CORPORATION | 4 |
| MICROSOFT CORPORATION | 6 | UNIVERSITY OF PENNSYLVANIA | 5 | DOLBY INTERNATIONAL | 5 | MICROSOFT TECH LICENSING | 4 |
| THOMSON LICENSING* | 5 | IBM | 3 | X-SYSTEM | 5 | SAMSUNG ELECTRONICS | 4 |
| | | | | | | MINAPSYS SOFTWARE CORP | 3 |
| | | | | | | MOTOROLA | 3 |

## 5. Conclusions

The frequency of patent citation to the scientific literature based on PATSTAT/Scopus is in the same order of magnitude as that based on Web of Science/USPTO, but slightly lower. It is as of yet uncertain whether this difference is due to differences in patent source (PATSTAT versus USPTO) or to differences in publication source (Scopus versus Web of Science). Large differences in patent-to-paper citations are found between disciplines, journal subject categories and individual journals. Research in biomedical fields tends to generate the largest impact upon technology, fully in agreement in findings in earlier studies such as those published by Francis Narin and his co-workers (e.g., Narin and Olivastro, 1992; Narin, Hamilton and Olivastro, 1997).

The study has shown that several fields and journals in social sciences do generate a considerable citation impact upon patentable technology. Counting citations in patents cited in one single year (2012 in the current study) to 1-5 year old journal papers, the journals in the subject categories Library & Information Science, Language & Linguistics, Education, and Law received more than 300 citations from patents. On the other hand, the patent-to-paper citation levels in humanities-related subject categories are extremely low, while those in social sciences tend to be higher, but lower than those obtained in biomedical fields. Also, the social science journals with the largest impact compared to other journals in the same category tend to show a strong orientation towards biomedical and natural science fields.

The journal citation impact measures calculated in the current paper are relatively simple, and fully compliant with the model underlying the journal impact factor proposed by Eugene Garfield and his collaborators. But more advanced indicators of technological impact of scientific journals could and should be calculated. The authors are currently carrying out a project aimed to develop more sophisticated indicators of the technological impact of scientific-scholarly journals.

The analysis of titles of citing patents and cited papers has given a useful impression of the topics involved in the patent-to-paper citation links, and thus, of the scientific-scholarly knowledge that is applied in patentable applications, as well as in the context in which it is used. As this type of analysis reaches beyond the level of sheer numbers, the current authors believe that it is important for a better understanding of what patent-to-paper citation counts actually measure, and thus may provide a basis for theoretical confidence in the use of patent citation statistics to study the science-technology interface. In order to be practically useful and convincing to users of patent citation-based impact indicators, it is therefore proposed to deliver not merely numbers, but also to provide information on the citation content, in terms of what is being cited, and in which context. World profiles of patent and paper titles presented in the current paper are useful tools but constitute a first step. More advanced mapping of citing and cited works is feasible and should be further explored. On the basis of the outcomes of such mapping studies, the theoretical understanding of what patent citations to SSH fields measure can be further expanded. Also, validation studies can address issue whether indicators based on these patent-paper citation links can be useful elements in the assessment of scholarly performance in SSH.